\documentclass[aps,superscriptaddress,twocolumn,amsmath,amssymb]{revtex4-1}
\usepackage{graphicx}
\usepackage{epstopdf}
\usepackage{bm}
\usepackage{subfigure}
\usepackage{sidecap}
\usepackage{times}
\usepackage{graphicx}
\usepackage{epstopdf}
\usepackage{amssymb}
\usepackage{amsmath}
\usepackage{dsfont,amsthm,amsbsy}
\usepackage{amssymb}
\usepackage{amsmath}
\usepackage{bbm}
\usepackage{graphicx}
\usepackage{epstopdf}
\usepackage{subfigure}
\usepackage{natbib}
\usepackage{epsfig}
\usepackage{amsfonts}
\usepackage{mathrsfs}
\usepackage{sidecap}
\usepackage{lipsum}
\usepackage{xcolor}
\usepackage[toc,page,title,titletoc,header]{appendix}
\usepackage[colorlinks,linkcolor=blue,citecolor=blue,anchorcolor=blue, urlcolor=blue]{hyperref}
\usepackage{hyperref}
\usepackage{resizegather}
\usepackage{float}
\usepackage{mathbbol}
\usepackage[normalem]{ulem}
\usepackage{cancel}
\usepackage{upgreek}
\usepackage{graphics,dcolumn,bm}
\usepackage{rotate,color}
\usepackage{times}
\usepackage{eqnarray}
\newcommand\redsout{\bgroup\markoverwith{\textcolor{red}{\rule[0.5ex]{2pt}{0.4pt}}}\ULon}

\newcommand{\bl}{\begin{aligned}}
\newcommand{\el}{\end{aligned}}
\def\be{\begin{equation}}
\def\ee{\end{equation}}

\def\bi{\begin{itemize}}
\def\ei{\end{itemize}}
\def\bn{\begin{enumerate}}
\def\en{\end{enumerate}}
\def\bea{\begin{eqnarray}}
\def\eea{\end{eqnarray}}
\def\no{\nonumber}
\def\ba{\begin{array}}
\def\ea{\end{array}}
\def\bd{\begin{displaymath}}
\def\ed{\end{displaymath}}

\graphicspath{{Figures/}}

\begin{document}

\title{Dissipative Floquet Dynamical Quantum Phase Transition}

\author{J. Naji}
\email[]{j.naji@ilam.ac.ir}
\affiliation{Department of Physics, Faculty of Science, Ilam University, Ilam, Iran}

\author{Masoud Jafari}
\email[]{masoud_jafari@comp.iust.ac.ir}
\affiliation{Department of Computer Engineering, Iran University of Science and Technology, Tehran, Iran}

\author{R. Jafari}
\email[]{jafari@iasbs.ac.ir, rohollah.jafari@gmail.com}
\affiliation{Department of Physics, Institute for Advanced Studies in Basic Sciences (IASBS), Zanjan 45137-66731, Iran}
\affiliation{Department of Physics, University of Gothenburg, SE 412 96 Gothenburg, Sweden}
\affiliation{Beijing Computational Science Research Center, Beijing 100094, China}

\author{Alireza Akbari}
\email[]{akbari@cpfs.mpg.de}
\affiliation{Max Planck Institute for the Chemical Physics of Solids, D-01187 Dresden, Germany}
\affiliation{Max Planck POSTECH Center for Complex Phase Materials,and Department of Physics, POSTECH, Pohang, Gyeongbuk 790-784, Korea}

\begin{abstract}
Non-Hermitian Hamiltonians provide a simple picture for inspecting dissipative systems with natural or in- duced gain and loss.
We investigate the Floquet dynamical phase transition in the dissipative periodically time driven XY and extended XY models, where the imaginary terms represent the physical gain and loss during the interacting processes with the environment.
The time-independent effective Floquet non-Hermitian Hamiltonians disclose three regions by analyzing the non-Hermitian gap: pure real gap (real eigenvalues), pure imaginary gap, and complex gap.
We show that each region of the system can be distinguished by the complex geometrical non-adiabatic phase.
We  have discovered that in the presence of dissipation, the Floquet dynamical phase transitions (FDPTs) still exist in the region where the time-independent effective Floquet non-Hermitian Hamiltonians reveal real eigenvalues.
Opposed to expectations based on earlier works on quenched systems, our findings show that the existence of the non-Hermitian topological phase is not an essential condition for dissipative FDPTs (DFDPTs).
We also demonstrate the range of driven frequency, over which the DFDPTs occur, narrows down by increasing the dissipation coupling and shrinks to a single point at the critical value of dissipation.
Moreover, quantization
and jumps of the dynamical  geometric phase reveals the topological characteristic feature of DFDPTs in the real gap region where confined to exceptional points.
\end{abstract}

\maketitle
\section{Introduction\label{Intro}}
Non-Hermitian Hamiltonians have recently attracted a lot of attention in the physics community across a wide range of fields, owing to their experimental feasibility~\cite{Hamazaki2021,li2019observation,Doppler2016,El2018,Lu2014topological,Chen2017,Xu2016,Feng2014,Hodaei2014,Hodaei2017,Peng2014loss,
Liertzer2012,Feng2013,Ozawa2019,Konotop2016,Ruter2010,Regensburger2012,Diehl2011}
and theoretical richness~\cite{Longhi2010,Carmichael1993,Lee2014,Malzard2015,Makris2008,Klaiman2008,
Lee2019,Ashida2018,Bender2007,Bender2005}.
Quantum systems driven by non-Hermitian Hamiltonians, display various fascinating
physical phenomena comparing to those governed by Hermitian Hamiltonians.
In cold-atom experiments, non-Hermitian Hamiltonians appear due to spontaneous decay~\cite{Dalibard1992,Plenio1998,Dum1992,Lee2014,Molmer1993}.
Furthermore, various non-Hermitian Hamiltonians have been utilized to treat various physical problems that might need to consider the interaction between the environment and the system, such as free-electron lasers~\cite{Dattoli1988}, topological lasers~\cite{Kartashov2019,Bandres2018,Harari2018}, electric circuit~\cite{Liu2020,Helbig2020,Hofmann2020}, transverse mode propagation in optical resonators~\cite{SIEGMAN1979}, multiphoton ionization~\cite{Baker1983}, many resonance phenomena~\cite{Moiseyev2011non}, nitrogen-vacancy-center in diamond~\cite{wu2019,Zhang2021},
with applications on high performance sensors~\cite{Wiersig2014,Lau2018,Hodaei2017,Chen2017}, and unidirectional transport devices~\cite{Lin2011,Feng2013}.

Theoretically, non-Hermitian Hamiltonians  trigger many novel physical
phenomena, such as non-Hermitian skin effect~\cite{Yao2018a,Yao2018b}, real eigenvalues with parity-time (PT) symmetry~\cite{Bender1998},
new topological properties corresponding to exceptional points (EPs)~\cite{Zhang2020,He2020,Shen2018,Bergholtz2021,Kawabata2019}, disorder induced self-energy in the effective Hamiltonian~\cite{Shen2018b,Papaj2019,Zhang2019,Tang2020,Jiang2019}, dynamical and topological properties~\cite{Bergholtz2021,Yuto2020,El2018,Konotop2016,Ygor2021}.
Finding the dynamical signatures of these non-equilibrium topological matter has become a fascinating area for more experimental and theoretical research.
 In recent works, several dynamical probes to the topological invariants of non-Hermitian phases in one and two dimensions have been introduced,
such as the non-Hermitian extension of dynamical winding numbers~\cite{Zhou2018,Longwen2021,Zhu2020,Zhou2018b,Zhou2019b,Zhou2021b}
and mean chiral displacements~\cite{Zhou2019,Pan2020}. Further, the dynamical quantum phase transitions (DQPTs)~\cite{Heyl2013,Jurcevic2017,Andraschko2014,Budich2016,Heyl2018,Abdi2019,Sedlmayr2018,Sedlmayr2018b,
Sedlmayr2020,Jafari2019,Bhattacharjee2018,Bhattacharya2017,Heyl2017,Guo2019,Tian2019,Wang2019,Uhrich2020,
Mishra2020,Jafari2019dynamical,Sadrzadeh2021,Yu2021,BozhenZhou2021,Porta2020,Sehrawat2021,Modak2021,Peotta2021,Juan,Kyaw2020,Lang2018}
following a quench across the EPs of a non-Hermitian lattice model is studied in Refs.~\cite{Longwen2021,Zhou2018}.
It has been shown that DQPTs appear for a quench from a trivial to a non-Hermitian topological phase~\cite{Longwen2021}.
This discovery indicates an underlying relationship between non-Hermitian topological phases and DQPTs.

To the best of our knowledge, the Floquet dynamical phase transition in the systems with gain and loss, and therefore
subject to non-unitary evolution, have not been addressed in prior publications and can provide a number of new insights into the subject.
This paper is devoted to the research on the Floquet dynamical phase transition~\cite{Kosior2018a,Kosior2018b,Qianqian2021,Zamani2020,Jafari2021,Yang2019} in the periodically time driven XY and extended XY spin models in the presence of dissipation.
The non-Hermitian terms (imaginary terms) represent dissipation--the physical gain and loss--when the chain interacts with the environment.
Our main purpose   is to study the effects of non-Hermitian terms on the FDPTs time and the range of driven frequency over which the DFDPTs occur.
First, we prob the phase diagram of the time-independent
effective Floquet non-Hermitian Hamiltonians by analyzing the energy gap of the systems analytically. We show that the phase diagram of
the system divided into three regions with pure real gap where confined to exceptional points, pure imaginary gap and complex gap.
We have found that, the region with real energy gap, where the DFDPTs occur,
is topologically nontrivial in the time-independent effective Floquet non-Hermitian XY Hamiltonian.
While the real gap region in the time-independent effective Floquet non-Hermitian extended XY Hamiltonian is topologically trivial.
In the other words, different from results obtained for the quenched case~\cite{Longwen2021}, existence of the non-Hermitian topologically nontrivial phase
is not necessary condition for appearance of the DFDPTs.
We have also shown that, the DFDPTs driven frequency range narrows down by increasing the dissipation coupling and shrinks to a single point at critical value of dissipation. We have found that adding the dissipation (imaginary term) to the Hermitian Hamiltonians affects those bounds of the driven frequency range
which correspond to the critical (gap closing) points of the time-independent effective Floquet Hermitian Hamiltonians.

\section{Dynamical phase transition\label{DQPT}}

The notion of a DQPT borrowed from the analogy between the partition function of an equilibrium system
$Z(\beta)={\rm Tr} [ e^{-\beta {\cal H}}]$ and the boundary quantum partition function $Z(z)=\langle\psi_{0}|e^{-z {\cal H}}|\psi_{0}\rangle$
with $|\psi_{0}\rangle$ a boundary state and $z \in \mathds{C}$.
When $z=it$, the boundary quantum partition function corresponds to a Loschmidt amplitude (LA),
${\cal L}(t)=\langle\psi_{0}|e^{-{\it i}  {\cal H}t}|\psi_{0}\rangle$,
expressing the overlap between the initial state $|\psi_{0}\rangle$
and the time-evolved one $|\psi_{0}(t)\rangle$~\cite{Heyl2013,Jurcevic2017,Andraschko2014,Budich2016,Heyl2018,Abdi2019,Yang2019,Sedlmayr2018,Sedlmayr2018b,
Sedlmayr2020,Jafari2019,Bhattacharjee2018,Bhattacharya2017,Heyl2017,Guo2019,Tian2019,Wang2019,Uhrich2020,
Zamani2020,Mishra2020,Jafari2019dynamical,Sadrzadeh2021,Jafari2021,Yu2021,BozhenZhou2021,Porta2020,Qianqian2021,Sehrawat2021,Modak2021}.
It has been argued that, like the thermal free energy, a dynamical free energy might well be defined as~\cite{Heyl2013}
%
%
\bea
\label{eq1}
g(t)=-\frac{1}{N}\lim_{N\rightarrow\infty} \ln|{\cal L}(t)|^{2}.
\eea
%
%
Here the real time $t$, plays the role of the control parameter and $N$ is the size of the system.
DQPTs are signaled by non-analytical behavior of dynamical free energy $g(t)$ as a function of time,
evincing in characteristic cusps in $g(t)$ or one of its time-derivatives~\cite{Heyl2013,Jurcevic2017,Andraschko2014,Budich2016,Heyl2018,Abdi2019,Yang2019,Sedlmayr2018,Sedlmayr2018b,
Sedlmayr2020,Jafari2019,Bhattacharjee2018,Bhattacharya2017,Heyl2017,Guo2019,Tian2019,Wang2019,Uhrich2020,
Zamani2020,Mishra2020,Jafari2019dynamical,Sadrzadeh2021,Jafari2021,Yu2021,BozhenZhou2021,Porta2020,Qianqian2021,Sehrawat2021,Modak2021}.
These cusps are followed by zeros of Loschmidt amplitude ${\cal L}(t)$, known in statistical physics as Fisher zeros of the
partition function~\cite{BozhenZhou2021,Heyl2018}.
Furthermore, analogous to order parameters at equilibrium quantum phase transition, a dynamical topological order parameter (DTOP)
is proposed to capture DQPTs~\cite{Budich2016}.
The DTOP is quantized and its unit magnitude jumps at the time of DQPT reveals the topological characteristic feature of DQPT~\cite{Budich2016,Bhattacharjee2018}.
This dynamical topological order parameter is extracted from the “gauge-invariant” Pancharatnam geometric
phase associated with the Loschmidt amplitude~\cite{Budich2016}.

The dynamical topological order parameter is defined as~\cite{Budich2016}
%
\begin{eqnarray}
\label{eq2}
\nu_D(t)=\frac{1}{2\pi}\int_0^\pi\frac{\partial\phi^G(k,t)}{\partial k}\mathrm{d}k,
\end{eqnarray}
%
where the geometric phase $\phi^G(k,t)$ is gained from the total phase $\phi(k,t)$ by subtracting the dynamical
phase $\phi^{D}(k,t)$:
$$
\phi^G(k,t)=\phi(k,t)-\phi^{D}(k,t).
$$
The total phase $\phi(k,t)$ is the phase factor of LA in its polar coordinates representation,
i.e., ${\cal L}_{k}(t)=|{\cal L}_{k}(t)|e^{i\phi(k,t)},$ results $\phi(k,t)=-i\ln\left[{\cal L}_{k}(t)/|{\cal L}_{k}(t)|\right]$, and
%
\bea
\bl
\label{eq3}
\phi^{D}(k,t)
= &
 -\int_{0}^{t}dt'\frac{\langle\psi_{-}(k,t')|
 {\cal H}_{k}(t')
 |\psi_{-}(k,t')\rangle}
{\langle\psi_{-}(k,t')|\psi_{-}(k,t')\rangle} \\
 &+
  \frac{i}{2}\ln\left[\frac{\langle\psi_{-}(k,t)|\psi_{-}(k,t)\rangle}{\langle\psi_{-}(k,0)|\psi_{-}(k,0)\rangle}\right].
\el
\eea
%
In following, to examine aspects of dissipative in quantum Floquet systems, we search for dissipative Floquet DPTs in proposed non-Hermitian periodically time driven Hamiltonians.

\section{Dissipative periodically time driven XY Model and Exact Solution\label{XYmodel}}
In this section we  study the phase diagram, topological properties and
FDPTs of dissipative periodically time driven XY model.
We show that the region in which DFDPTs occur is confined to exceptional points and
is topologically nontrivial and the time-independent effective Floquet non-Hermitian Hamiltonian
has real eigenvalues.

\subsection{Exact solution\label{ESXY}}

The Hamiltonian of $N$ sites dissipative periodically time driven XY spin model is given as
%
\begin{equation}
\bl
\label{eq4}
{\cal H}(t)
\!
=
\!\!
\sum_{n}
&
\Big[
[J
\!
-
\!
\gamma\cos(\omega t)]S_{n}^{x}S_{n+1}^{x}
+
[J
\!
+
\!
\gamma\cos(\omega t)]S_{n}^{y}S_{n+1}^{y}
\\
&-\gamma\sin(\omega t)(S_{n}^{x}S_{n+1}^{y}+S_{n}^{y}S_{n+1}^{x})+h S_{n}^{z}
\\
&-
{\it i}(\Gamma_{u}S_{n}^{+}S_{n}^{-}+\Gamma_{d}S_{n}^{-}S_{n}^{+})\Big],
\el
\end{equation}
%
where $S_{n}^{\alpha=\{x,y,z\} }=\sigma^{\alpha}/2$, 
and $\sigma^{\alpha}$ are Pauli matrices.
Furthermore,
$S_{n}^{\pm}=\sigma^{\pm}/2=(\sigma^{x}\pm {\it i}\sigma^{y})/2$ are the spin raising and lowering operators
which correspond to the gain $\Gamma_{u}<0$ ($\Gamma_{d}<0$) or loss $\Gamma_{u}>0$ ($\Gamma_{d}>0$) of
spin up state $|\uparrow\rangle$ (spin down state $|\downarrow\rangle$) during the interacting processes with the environment
with the rate of $\Gamma_{u}$ ($\Gamma_{d}$), and $\omega$ is the driving frequency.
The system can be reduced to the Floquet Hermitian XY model when $\Gamma_{u}=\Gamma_{d}=0$~\cite{Yang2019}.
The term ``dissipative" refers to the system's tunneling effects to its own continuum, which is common in quantum optics and nuclear physics when using the Feshbach projection method on intrinsic states.

The Hamiltonian, Eq.~(\ref{eq1}),  can be mapped to the free spinless fermion model with complex
chemical potential~\cite{Zeng2016} by Jordan-Wigner transformation~\cite{LIEB1961,Barouch1971,Jafari2011,Jafari2012} (see Appendix \ref{AA})
%
\bea
\bl
\label{eq5}
{\cal H}(t)= \sum_{n=1}^{N}
\Big[
&
\Big(\frac{J}{2} c_{n}^{\dagger} c_{n+1}
-\frac{\gamma}{2} e^{-{\it i} \omega t} c_{n}^{\dagger} c^{\dagger}_{n+1}+
{\rm H.C.}
\Big)
\quad
\\
&+
(h-{\it i}\Gamma_{-}) c_{n}^{\dagger} c_{n}-{\it i}\Gamma_{+}\Big],
\el
\eea
%
where
$\Gamma_{\pm}=\Gamma_{u} \pm \Gamma_{d}$,
and
$c_{n}^{\dagger}$ ($c_{n}$) are
the spinless fermion creation (annihilation) operators, respectively.
Thanks to the Fourier transform, the Hamiltonian ${\cal H}(t)$ in Eq.~(\ref{eq5})
can be written as the sum of $N/2$ non-interacting terms
$${\cal H}(t) = \sum_{k>0} {\cal H}_{k}(t)$$
where ${\cal H}_{k}(t)=C^{\dagger}\mathbb{H}_{k}(t)C-{\it i}\Gamma_{+}\mathbb{1}$
with
$C^{\dagger}=(c_{k}^{\dagger},~c_{-k})$, and
%
\bea
\bl
\label{eq6}
\mathbb{H}_{k}(t)=
\left(
\begin{array}{cc}
h_{z}(k) & {\it i}h_{xy}(k)e^{-{\it i} \omega t} \\
-{\it i}h_{xy}(k)e^{{\it i} \omega t}  & -h_{z}(k) \\
\end{array}
\right).
\el
\eea
%
 The parameters $h_{xy}(k)$ and $h_{z}(k) $ are given as $h_{xy}(k)=\gamma\sin(k)$, and $h_{z}(k)=J\cos(k)+h-{\it i}\Gamma_{-}$.
Using the time-dependent Schr\"{o}dinger equation
${\it i}\frac{d}{dt}|\psi_{k}^{\pm}(t)\rangle=\mathbb{H}_{k}(t)|\psi_{k}^{\pm}(t)\rangle$ in
the rotating frame given by the non-unitary transformation
$U(t)=U_{R}(t)U_{D}(t)$, with $U_{R}(t)=\exp[{\it i}\omega(\mathbb{1}-\sigma^{z})t/2]$, and
$U_D(t)= e^{-\Gamma_{+}t}\mathbb{1}$, the time-dependent Hamiltonian is transformed to the time-independent
effective Floquet non-Hermitian form (see Appendix \ref{AA})
%
\bea
\label{eq7}
H_{F}(k)=-h_{xy}(k)\sigma_{y}+\Big(h_{z}(k)-\frac{\omega}{2}\Big)\sigma_{z}+\frac{\omega}{2}\mathbb{1}.
\eea
%
Then the time-evolved $|\psi_{k}(t)\rangle$ of the quasi-spin Hamiltonian $H_{k}(t)$, is given by
%
{\small
\begin{eqnarray}
|\psi_{k}(t)\rangle&=&U(t)e^{-iH_{F}(k)t}|\varphi_{k}\rangle,
\label{eq8}
\end{eqnarray}
}
%
where $|\varphi_{k}\rangle$ is the initial state of the system at $t=0$.
Due to the decoupling of different momentum sectors, the initial and time-evolved ground states of
the original Hamiltonian exhibit a factorization property that is expressed by
%
\begin{equation}
\bl
&
|\psi(t)\rangle
=
\prod_{k}|\psi_{k}(t)\rangle=\prod_{k} U(t)e^{-iH_{F}(k)t}|\varphi_{k}\rangle,
\\
&
|\psi(t=0)\rangle
=
\prod_{k}|\varphi_{k}\rangle.
\label{eq9}
\el
\end{equation}
%
We consider that at $t=0$ the system prepared at $|\psi(0)\rangle=|\varphi_{k}\rangle=|\downarrow\rangle$, i.e.,
$c_1(t=0)=0$ and $c_2(t=0)=1$, where $c_1$ and $c_2$ are probability amplitudes of
$|\psi(0)\rangle$ at up ($|\uparrow\rangle$) and down ($|\downarrow\rangle$) states, respectively.
Then according to Eq.~(\ref{eq9}) the unnormalized time evolving state $|\psi(k,t)\rangle$ of the Hamiltonian ${\cal H}_{k}(t)$ is given by:
%
\bea
\label{eq10}
\bl
|\psi(t)\rangle
=&
\prod_{k}|\psi(k,t)\rangle,
\el
\eea
with
\bea
\bl
&
|\psi(k,t)\rangle
=
\Big[e^{-\Gamma_{+} t}\Big(\frac{h_{xy}(k)}{\Lambda}\sin(\Lambda t)\Big)|\uparrow\rangle
\\
&\quad
+
e^{-\Gamma_{+} t} e^{{\it i}\omega t}
\Big(
\!\!
\cos(\Lambda t)+
{\it i}\frac{2h_{z}(k)-\omega}{2\Lambda}
\sin(\Lambda t)\Big)|\downarrow\rangle\Big],
\;\;\quad
\el
\eea
%
and
$\Lambda=\sqrt{h_{xy}^{2}(k)+[h_{z}(k)-\frac{\omega}{2}]^{2}}$.

The time-independent effective Floquet non-Hermitian Hamiltonian in Eq.~(\ref{eq7}), possesses the sublattice symmetry
$U_{s}H(k)U^{-1}_{s}=-H(k)$ with ${\cal S}=\sigma_{x}$,
and generalized particle-hole symmetry
$U_{p}H^{\top}(k)U^{-1}_{p}=-H(-k)$,
as well as, the time-reversal symmetry
$U_{T}H^{\top}(k)U^{-1}_{T}=H(-k)$ with  $U_{p}=\sigma_{x}$ and $U_{T}=\mathbb{1}$.
Here $H^{\top}(k)$ is transposed of $H^{\top}(k)$. Consequently, the symmetry class of the non-Hermitian
time independent Hamiltonian in Eq.~(\ref{eq7}) belongs to BDI in the periodic table of non-Hermitian
topological phases~\cite{Kawabata2019}.
Moreover, $H_{F}(k)$ encompasses the inversion symmetry $U_{I}H(k)U^{-1}_{I}=H(-k)$ with $U_{I}=\sigma_{z}$, which
manifests the correspondence between the bulk topological invariant and the number of Majorana edge modes under
the open boundary condition~\cite{Kawabata2019,Zeng2016}.

\begin{figure}[t!]
\centerline{\includegraphics[width=\columnwidth]{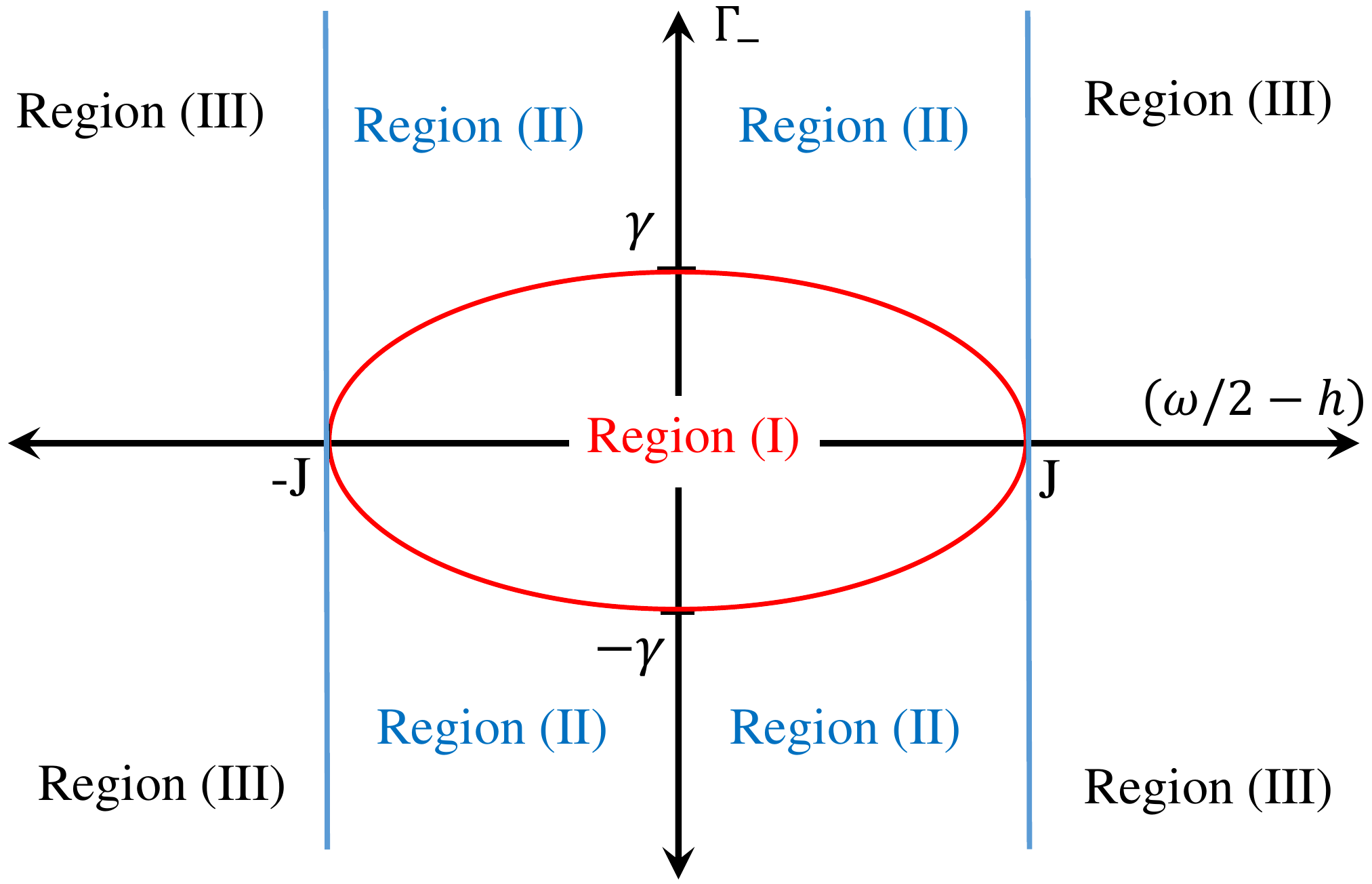}}
\centering
\caption{(Color online) The Phase diagram of the time independent
effective Floquet non-Hermitian XY Hamiltonian. The
red line denotes the exceptional ring, which corresponds elliptic
equation (Eq.~(\ref{eq13})). In region~(I), the system is in the
ferromagnetic phase with a pure real energy gap. In region~(II) the energy
gap of the non-Hermitian Hamiltonian is a pure imaginary. In region~(III)
it is in the paramagnetic phase with a complex non-Hermitian gap.}
\label{fig1}
\end{figure}

The complex energy spectrum of $H_{F}$ is given as
%
\begin{equation}
\no
\varepsilon^{\pm}_{k}=\frac{\omega}{2}\pm\sqrt{h_{xy}^{2}(k)+[h_{z}(k)-\frac{\omega}{2}]^{2}}.
\end{equation}
%
and becomes gapless if
%
\bea
\bl
\label{eq11}
&2\Gamma_{-}[J\cos(k)+h-\frac{\omega}{2}]=0,\\
&[J\cos(k)+h-\frac{\omega}{2}]^{2}+
[\gamma\sin(k)]^{2}-
\Gamma_{-}^{2}=0.
\el
\eea
%
By solving the above equations, we can get
%
\bea
\bl
\label{eq12}
&k^{\ast}=\arccos(\frac{\omega-2h}{2J}),\\
\label{eq13}
&\frac{\Gamma_{-}^2}{\gamma^2}+\frac{(\omega-2h)^2}{4J^2}=1.
\el
\eea
%
%
The Eq.~(\ref{eq12}) implies a limitation $\omega-2h<\pm 2J$ and the Eq.~(\ref{eq14}) depicts an elliptical
exceptional ring. Therefore, the system can be separated into three regions as shown in Fig.~\ref{fig1}.
%
%
In the region~(I), inside the exceptional ring, the energy gap (eigenvalues) $\Delta=|\varepsilon_{K}^{+}-\varepsilon_{k}^{-}|$
is purely real i.e., ${\rm Im}[\Delta]=0$ and  Re$[\Delta]>0$
(Im[$\mathbb{C}$] and Re[$\mathbb{C}$] represent the imaginary and real part of complex
number $\mathbb{C}$, respectively).
In this region $k^{\ast}=\arccos[(\omega-2h)/(2J)]$
and
$\Delta=\sqrt{\gamma^{2}[1-(\omega-2h)^2/(4J^{2})]-\Gamma_{-}^{2}}$.
In the region~(II) the gap is pure imaginary, i.e., Im$[\Delta]\neq0$ and Re$[\Delta]=0$.
In this region we still have $k^{\ast}=\arccos[(\omega-2h)/(2J)]$ but
the non-Hermitian strength $\Gamma_{-}$ is large enough to be dominant, then
$\Delta={\it i}\sqrt{\Gamma_{-}^{2}-\gamma^{2}[1-(\omega-2h)^2/(4J^{2})]}$.
The region~(III) ($|\omega-2h|>2J$) is characterised by the complex gap. In the other words, in the region~(III)
both real and imaginary parts of the gap is non-zero.

\subsection{Complex geometrical non-adiabatic phase\label{GPXY}}
In this section we  study the geometric phase of the model to show how the
geometric phase can detect the three regions in the time-independent effective Floquet non-Hermitian
XY Hamiltonian mentioned in previous section.
For the non-adiabatic evolutions we use the Lewis-Riensenfeld invariant theory~\cite{Lewis1969} which generalized
to non-Hermitian systems~\cite{Gao1992,GARRISON1988}.
%
\begin{figure*}[t!]
\centerline{\includegraphics[width=\linewidth]{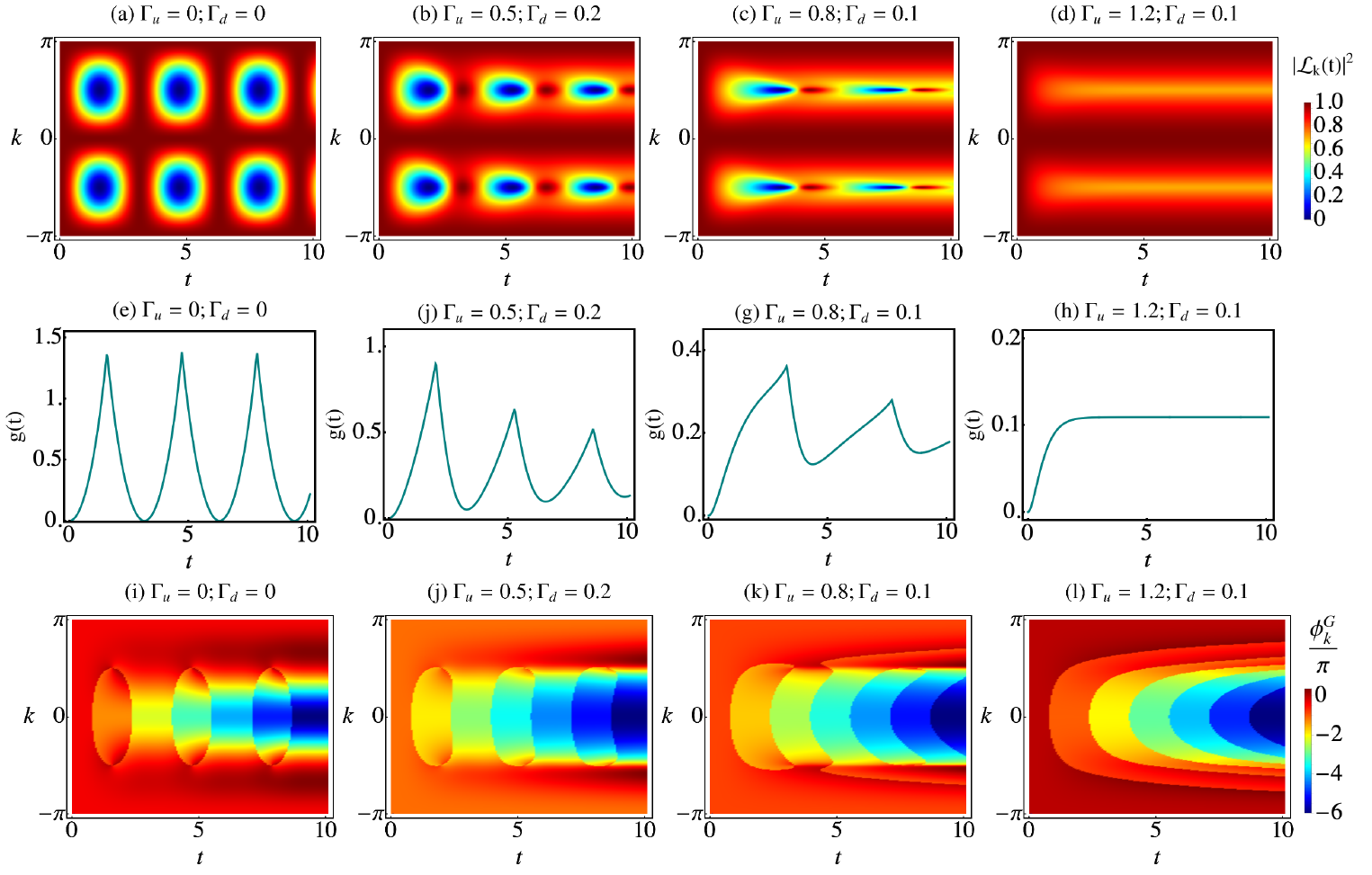}}
\centering
\caption{(Color online) The density plot of Loschmidt echo $|{\cal L}_{k}(t)|^{2}$
of periodically time driven XY model as a function of time $t$ and $k$, for  (a) $\Gamma_{-}=0$,
(b) $\Gamma_{-}=0.3$, (c) $\Gamma_{-}=0.7$, (d) $\Gamma_{-}=1.1$.
The dynamical free energy of the model versus time $t$ for (e) $\Gamma_{-}=0$,
(f)  $\Gamma_{-}=0.3$, (g) $\Gamma_{-}=0.7$, (g) $\Gamma_{-}=1.1$.
The density plot of geometric phase as a function of time and $k$
for  (i) $\Gamma_{-}=0$, (j) $\Gamma_{-}=0.3$, (k) $\Gamma_{-}=0.7$,
(l) $\Gamma_{-}=1.1$.
In all plots we set $J=h=\gamma=1$ and $\omega=2$.
}
\label{fig2}
\end{figure*}
%
According to Lewis-Riensenfeld theory the non-Hermitian invariant $I(t)$ associated
to the Hamiltonian $\mathbb{H}_{k}(t)$, Eq.~(\ref{eq6}), can be expressed as linear
combinations of Pauli matrices, i.e.,

%
\bea
\label{eq14}
I(t)=r_{1}S^{+}+r_{2}(t)S^{-}+r_{3}(t)S^{z}.
\eea
%
where $r_{m=\{1,2,3\}}(t)$ are three time-dependent complex parameters and $I(t)$ satisfies the Liouville-von Neumann equation
%
\bea
\label{eq15}
\frac{d}{dt}I(t)=\frac{\partial}{\partial t}I(t)-{\it i}\Big[I(t),\mathbb{H}_{k}(t)\Big].
\eea
%
The substitution of expressions of $I(t)$ and $\mathbb{H}_{k}(t)$ in Eq.~(\ref{eq15}) leads to the
system of coupled differential equations. By solving the coupled differential equations, which satisfies the cyclicity of $I(t+T)=I(t)$
with $T=2\pi/\omega$ results
%
\bea
\label{eq16}
I(t)=
\left(
\begin{array}{cc}
\cos(\alpha) & \sin(\alpha)e^{-{\it i}\omega t} \\
\sin(\alpha)e^{{\it i}\omega t} &  -\cos(\alpha) \\
\end{array}
\right),
\eea
%
where $\cos(\alpha)=\frac{2h_{z}(k)-\omega}{\sqrt{4h_{xy}^{2}(k)+[2h_{z}(k)-\omega]^{2}}}$.

The complex geometrical non-adiabatic phase for a cyclic evolution $T=2\pi/\omega$ is defined by~\cite{Gao1992,GARRISON1988}
%
\bea
\label{eq17}
\beta(t)={\it i}\int_{0}^{T}\langle\Phi_{-}|\frac{\partial}{\partial t}|\Psi_{-}\rangle dt,
\eea
%
where $|\Psi_{-}\rangle$ and $|\Phi_{-}\rangle$, are the instantaneous eigenstates of $I(t)$ and $I(t)^{\dag}$ (see Appendix \ref{AB}).
The complex geometrical non-adiabatic phase for the periodically time driven dissipative Floquet XY model
is obtained as
%
\bea
\bl
\no
\label{eq18}
\beta=\pi
[1-\cos(\alpha)]
=
\pi
\Big[
1- 
\frac{2h_{z}(k)-\omega}{\sqrt{4h_{xy}^{2}(k)+[2h_{z}(k)-\omega]^{2}}}
\Big],
\el
\\
\eea
%
which is a generalization of the complex solid angle in complex parameter space~\cite{GARRISON1988}.
The real part of the complex geometrical non-adiabatic phase is given by
%
\bea
\no
{\rm Re}[\beta]=
\left\{
\begin{array}{ll}
\pi, & \hbox{\small\text{Region(I)}} \\
\pi
\Big[
1+\frac{\Gamma_{-}}{\sqrt{\Gamma_{-}^{2}-\gamma^{2}[1-(\omega-2h)^{2}/(4J^{2})]}}
\Big], & \hbox{\small\text{Region(II)}} \\
\pi
\Big[1-f(k)
\Big]
, & \hbox{\small\text{Region(III)}}
\end{array}
\right.
\eea
%
where
$$
f(k)=
\frac{
2h+2J\cos(k)-{\omega}
}{
2{\rm Re}[\Delta]
}
-
\frac{
\Gamma_{-}
}{
{\rm Im}[\Delta]
}
.
$$
As seen the real part of the complex geometrical non-adiabatic phase shows singularity at
phase boundaries. In addition, the real part of complex geometrical non-adiabatic phase
in region~(I), which confined to exceptional points, is independent of the Hamiltonian parameters.
In the next section we will study the topological properties of the effective Hamiltonian in Eq.~(\ref{eq7})
using the winding numbers of the non-Hermitian Hamiltonians~\cite{Zhu2020}.

\subsection{Topological Invariant\label{TIXY}}
Examining the non-Hermitian Hamiltonians' winding numbers expresses that  both inside and outside the exceptional ring is distinguished by its winding number,
$N_w=\frac{1}{2\pi}\int_{-\pi}^{\pi}\partial_{k}\phi(k)dk$.
Here
 $\phi(k)=\arctan\left[(\omega- 2h_{z}(k))/(2h_{y}(k))\right]$
is winding angle~\cite{Zhu2020}. The winding numbers of the non-Hermitian topological and trivial phases are found to be $N_w=1$
for
$(\Gamma_{-}^2/\gamma^2)+[(\omega-2h)^2/(4J^2)]<1$
(inside the exceptional ring) and $N_w=0$ for outside the exceptional ring, respectively.
As can be seen, the topological phase spreads by dissipation which is
unique to non-Hermitian systems.
%
%

\subsection{Pure state dynamical topological quantum phase transition\label{PDPTXY}}
As obtained in Eq.~(\ref{eq10}), if at $t=0$ the system prepared at $|\psi_{-}(0)\rangle=|\downarrow\rangle$,
the unnormalized time evolved initial state of the dissipative Floquet XY Hamiltonian is expressed as:
%
\bea
\bl
\no
\label{eq19}
&
|\psi_{-}(k,t)\rangle
=
\Big[e^{-\Gamma_{+} t}\Big(\frac{h_{xy}(k)}{\Lambda}\sin(\Lambda t)\Big)|\uparrow\rangle
\\
&
\quad
+
e^{-\Gamma_{+} t} e^{{\it i}\omega t}\Big(\cos(\Lambda t)
+
{\it i}
\frac{2h_{z}(k)-\omega}{2\Lambda}\sin(\Lambda t)\Big)|\downarrow\rangle\Big].
\el
\\
\eea
%
It is straightforward to see how the return probability (LA) is determined
%
\begin{equation}
\bl
\label{eq20}
{\cal L}(k,t)=e^{-\Gamma_{+} t}e^{{\it i}\omega t}
\Big[
\frac{\cos(\Lambda t)+
{\it i}
\frac{2h_{z}(k)-\omega}{2\Lambda}\sin(\Lambda t)}
{\sqrt{\langle\psi_{-}(k,t)|\psi_{-}(k,t)\rangle}}
\Big].
\el
\end{equation}
%
%
The FDQPT occurs at the time instances at which at least one factor in LA becomes zero i.e., ${\cal L}_{k^{\ast}}(t^{\ast})=0$
which yields
%
\bea
\label{eq21}
t^{\ast}=\frac{-{\it i}}{2\Lambda}\ln\Big[
\frac{2h_{z}-\omega -2\Lambda}
{2h_{z}-\omega+2\Lambda}
\Big].
\eea
%
By a rather lengthy calculation, one can obtain that there are real solutions of $t^{\ast}$ only whenever
%
\bea
\label{eq22}
2
\Big(
h-J\sqrt{1-\frac{\Gamma_{-}^2}{\gamma^2}}
\Big)
<\omega<
2
\Big(
h+J\sqrt{1-\frac{\Gamma_{-}^2}{\gamma^2}}
\Big),
\eea
%
at quasi-momentum $k^{\ast}=\arccos[(\omega-2h)/(2J)]$ results
%
\bea
\label{eq23}
t^{\ast}=\frac{1}{2\Lambda}(2n+1)\pi+\frac{1}{\Lambda}\arctan(\frac{\Gamma_{-}}{\Lambda}).
\eea
%
DFDPTs arise in the range of driving frequency over which the eigenvalues of the time-independent effective Floquet non-Hermitian XY Hamiltonian are purely real and the system is also topological, since Eq.~(\ref{eq22}) is nothing but Eq.~(\ref{eq13}).
On the other hand, when
$\Gamma_{-}^2/\gamma^2+(\omega-2h)^2/(4J^2)>1$, there is no critical momentum and $t^{\ast}$ is always complex
resulting no DFDPTs at any given real time $t$.
We should note that, the term $(2n+1)\pi/(2\Lambda)$ in Eq.~(\ref{eq23})
is the FDPTs time scale in the absence of dissipation and the term $[\arctan(\Gamma_{-}/\Lambda)]/\Lambda$
originates from the dissipation. As is clear, both the lower and upper bounds of the range of driven frequency
over which DFDPTs occur are function of the dissipation.
Thus, the DFDPT driven frequency range shrinks to a single point
$\omega=2h$ at $\Gamma_{-}=\pm\gamma$.
When the gain or loss of the spin up and down states are equal,  $\Gamma_{u}=\Gamma_{d}\neq0$, the DFDPT times drop to non-dissipative FDPT times even in the presence of dissipation, as shown by Eq.~(\ref{eq23}).
%
In such a case, the system is in the resonance regime where the population
completely cycles the population between the two spin down and up states.
It is worthwhile to mention that, the anisotropy $\gamma$ does not affect the non-dissipative FDPTs driven frequency
range ($\Gamma_{u}=\Gamma_{d}=0$)~\cite{Zamani2020,Jafari2021,Yang2019}, while the DFDPTs driven frequency
range controls by $\gamma$.

The numerical simulation of the density plot of the Loschmidt echo $|{\cal L}(k,t)|^{2}$, the dynamical free
energy $g(t)$ and density plot of the geometric phase have been depicted in Fig.~\ref{fig2} for the Hamiltonian
parameters inside and out side the exceptional ring.
When the time-independent effective non-Hermitian XY Hamiltonian $H_F$ is in non-Hermitian topological phase, it is apparent that
there exist critical points $k^{\ast}$ and $t^{\ast}$, where ${\cal L}_{k^{\ast}}(t^{\ast})$ becomes zero [Figs.~\ref{fig2}(a)-\ref{fig2}(c)].
Outside of the exceptional ring, however, there is no such critical point [Fig.~\ref{fig2}(d)].
Moreover, in Figs.~\ref{fig2}(e)-\ref{fig2}(g) the DFDPTs are observed as the cusps in $g(t)$ for the driving frequency at which the system inters into
the non-Hermitian topological phase. While the dynamical free energy shows completely analytic, smooth behavior for the Hamiltonian parameters set out side
the exceptional ring.

The density plot of geometric phase are plotted in Figs.~\ref{fig2}(i)-\ref{fig2}(l) for different values of Hamiltonian's parameters
inside and out side of the exceptional ring.
As seen, the plots display singular changes at critical times $t^{\ast}$, and
at critical momentum $k^{\ast}$ when the system is in region~(I), while it shows smooth behavior for the case that the DPTs are absent.
This behaviour represents the topological aspects of DFDPTs, where the phase of the time-independent effective Floquet non-Hermitian
XY Hamiltonian is topological.

\section{Dissipative periodically time driven extended XY Model\label{EXYmodel}}
In this section we  study the phase diagram, topological properties and
FDPTs of dissipative periodically time driven extended XY (EXY) model.
We  show that the region where
%
\begin{figure}[t!]
\centerline{\includegraphics[width=\columnwidth]{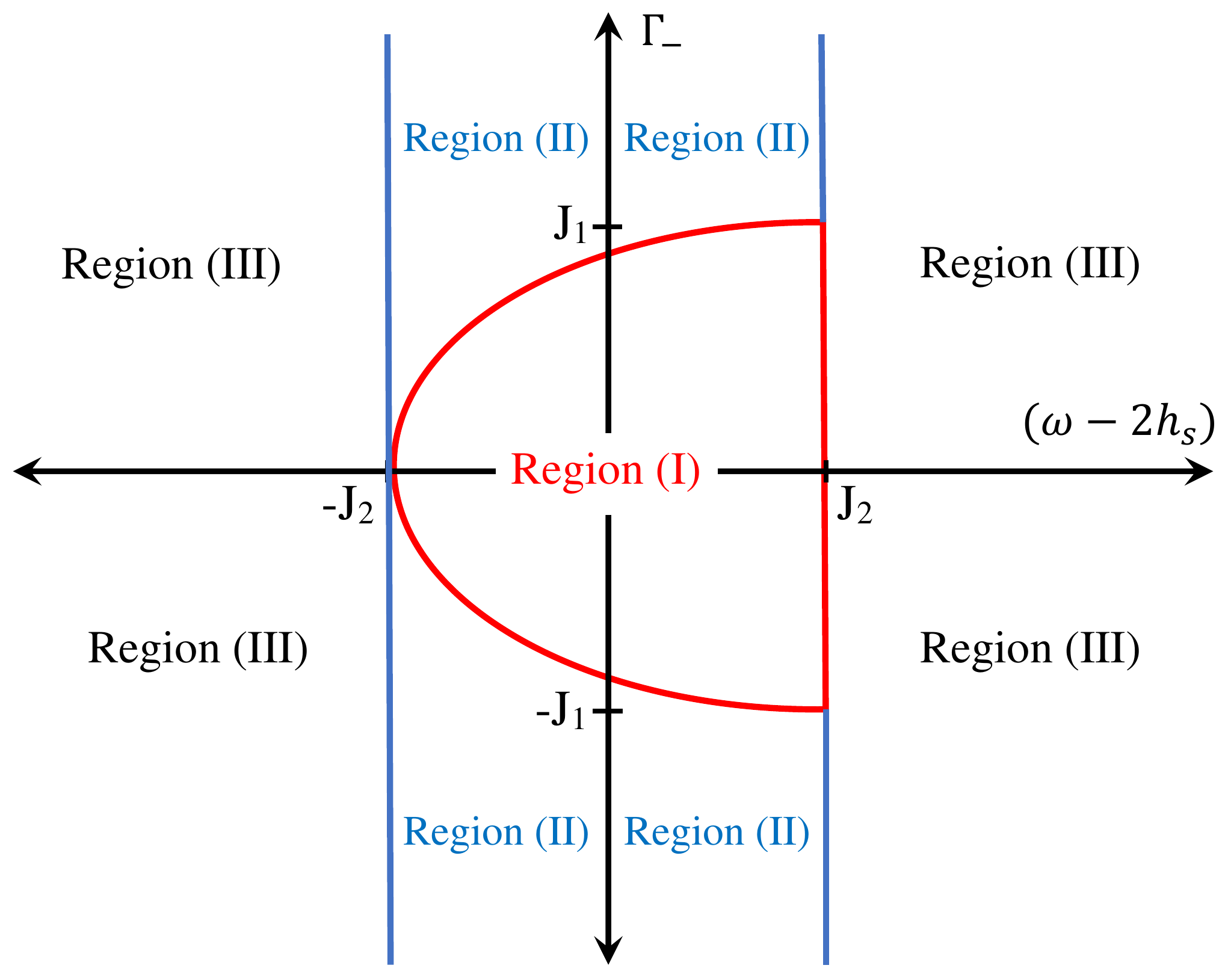}}
\centering
\caption{(Color online) The Phase diagram of the time independent
effective Floquet non-Hermitian extended XY Hamiltonian. The
red line denotes the exceptional ring, which corresponds to
Eq.~(\ref{eq29}). In region~(I), the eigenvalues (gap) of the time-independent
effective Floquet non-Hermitian extended XY Hamiltonian is
purely real . In region~(II) the energy gap of the effective non-Hermitian
Hamiltonian is a pure imaginary. In region~(III) the energy gap
is complex.}
\label{fig3}
\end{figure}
%
DFDPTs occur is confined to exceptional points and the time-independent effective Floquet non-Hermitian EXY Hamiltonian
has real eigenvalues but the system is topologically trivial.

\subsection{Exact solution\label{ESEXY}}

The Hamiltonian of the one-dimensional harmonically driven extended XY spin chain in the staggered magnetic field is given by~\cite{Zamani2020}
%
\begin{eqnarray}
\bl
\label{eq24}
{\cal H}(t) = \sum_{n=1}^{N}
&
\Big[J_{1}\cos(\omega t) \Big(S_n^x S_{n+1}^x + S_n^y S_{n+1}^y \Big),\\
&- (-1)^{n} J_{1} \sin(\omega t) \Big(S_n^x S_{n+1}^y - S_n^y S_{n+1}^x \Big)\\
\nonumber
&-(-1)^{n} J_{2} \Big(S_n^x S_{n+1}^z S_{n+2}^x + S_n^y S_{n+1}^z S_{n+2}^y \Big)\\
&-{\it i}(\Gamma_{u}S_{n}^{+}S_{n}^{-}+\Gamma_{d}S_{n}^{-}S_{n}^{+})+(-1)^{n} h_{s} S_n^z \Big].
\el
\\
\end{eqnarray}
%
The first and second terms in Eq.~(\ref{eq24}) describe the time dependent nearest neighbour XY and staggered
Dzyaloshinskii-Moriya interactions~\cite{Jafari2011b}, and the third term is a staggered cluster (three-spin)
interaction~\cite{Titvinidze}.

This Hamiltonian 
can be exactly diagonalized by Jordan-Wigner transformation~\cite{LIEB1961,Barouch1971,Jafari2011,Jafari2012}
which transforms spins into spinless fermions, where $c^{\dagger}_{n}$ ($c_{n}$) is the fermion creation (annihilation)
operator~\cite{Zamani2020}. The crucial step is to define two independent fermions at site $n$, $c_{n-1/2}^{A}=c_{2n-1}$, and $c_{n}^{B}=c_{2n}$,
which can be regarded as splitting the chain having a diatomic unit cell. The Fourier transformed Hamiltonian can be expressed
as sum of independent terms ${\cal H}(t)=\sum_{k}{\cal H}_{k}(t)$ with ${\cal H}_{k}(t)=\Psi^{\dagger}\mathbb{H}_{k}(t)\Psi-{\it i}\Gamma_{+}\mathbb{1}$ and $\Psi^{\dagger}_k=(c_{k}^{\dagger B},~c_{k}^{\dagger A})$,
where the Bloch single particle Hamiltonian $\mathbb{H}_{k}(t)$ is given as  $\mathbb{H}_{k}(t)=[h_{xy}(\cos(\omega t)\sigma^{x}+\sin(\omega t)\sigma^{y})+h_{z}\sigma^{z}]$,
with $h_{xy}(k)=J_{1}\cos(k/2)$ and $h_{z}(k)=J_{2}\cos(k)/2+h_{s}-{\it i}\Gamma_{-}$.

Using the time-dependent Schr\"{o}dinger equation ${\it i}\frac{d}{dt}|\psi_{k}^{\pm}(t)\rangle=\mathbb{H}_{k}(t)|\psi_{k}^{\pm}(t)\rangle$ in
the rotating frame given by the periodic non-unitary transformation $U(t)=U_{R}(t)U_{D}(t)$, with
$U_{R}(t)=\exp[{\it i}\omega(\mathbb{1}-\sigma^{z})t/2]$, and $U_D(t)= e^{-\Gamma_{+}t}\mathbb{1}$
the time dependent Hamiltonian is transformed to the time-independent effective Floquet non-Hermitian form
%
\begin{equation}
H_F(k)=h_{xy}(k)\sigma^{x}+(h_{z}(k)-\frac{\omega}{2})\sigma^{z}+\frac{\omega}{2}\mathbb{1}
\label{eq25}.
\end{equation}
%
Following the calculation in section~\ref{ESXY}, if 
 at $t=0$ the system prepared at
$|\psi(0)\rangle=|\varphi_{k}\rangle=|\downarrow\rangle$, then according to Eqs.~(\ref{eq8})~and~(\ref{eq9})
the unnormalized time evolving state $|\psi(k,t)\rangle$ of the Hamiltonian ${\cal H}_{k}(t)$ is given by:
%
{
\bea
\bl
\no
\label{eq26}
&
|\psi(t)\rangle=
\prod_{k}|\psi(k,t)\rangle,\\
&
|\psi(k,t)\rangle=
\Big[e^{-\Gamma_{+} t}\Big(-{\it }i\frac{h_{xy}(k)}{\Lambda}\sin(\Lambda t)\Big)|\uparrow\rangle
\\
&
\quad
+
e^{-\Gamma_{+} t} e^{{\it i}\omega t}\Big(\cos(\Lambda t)+
{\it i}
\frac{2h_{z}(k)-\omega}{2\Lambda}\sin(\Lambda t)\Big)|\downarrow\rangle\Big],
\el
\\
\eea
}
%
with $\Lambda=\sqrt{h_{xy}^{2}(k)+[h_{z}(k)-\frac{\omega}{2}]^{2}}$.

The complex energy spectrum of $H_{F}$ is given as
%
\begin{equation}
\no
\epsilon^{\pm}_{k}=\frac{\omega}{2}\pm\sqrt{h_{xy}^{2}(k)+[h_{z}(k)-\frac{\omega}{2}]^{2}},
\end{equation}
%
%
\begin{figure*}[ht!]
\centerline{\includegraphics[width=\linewidth]{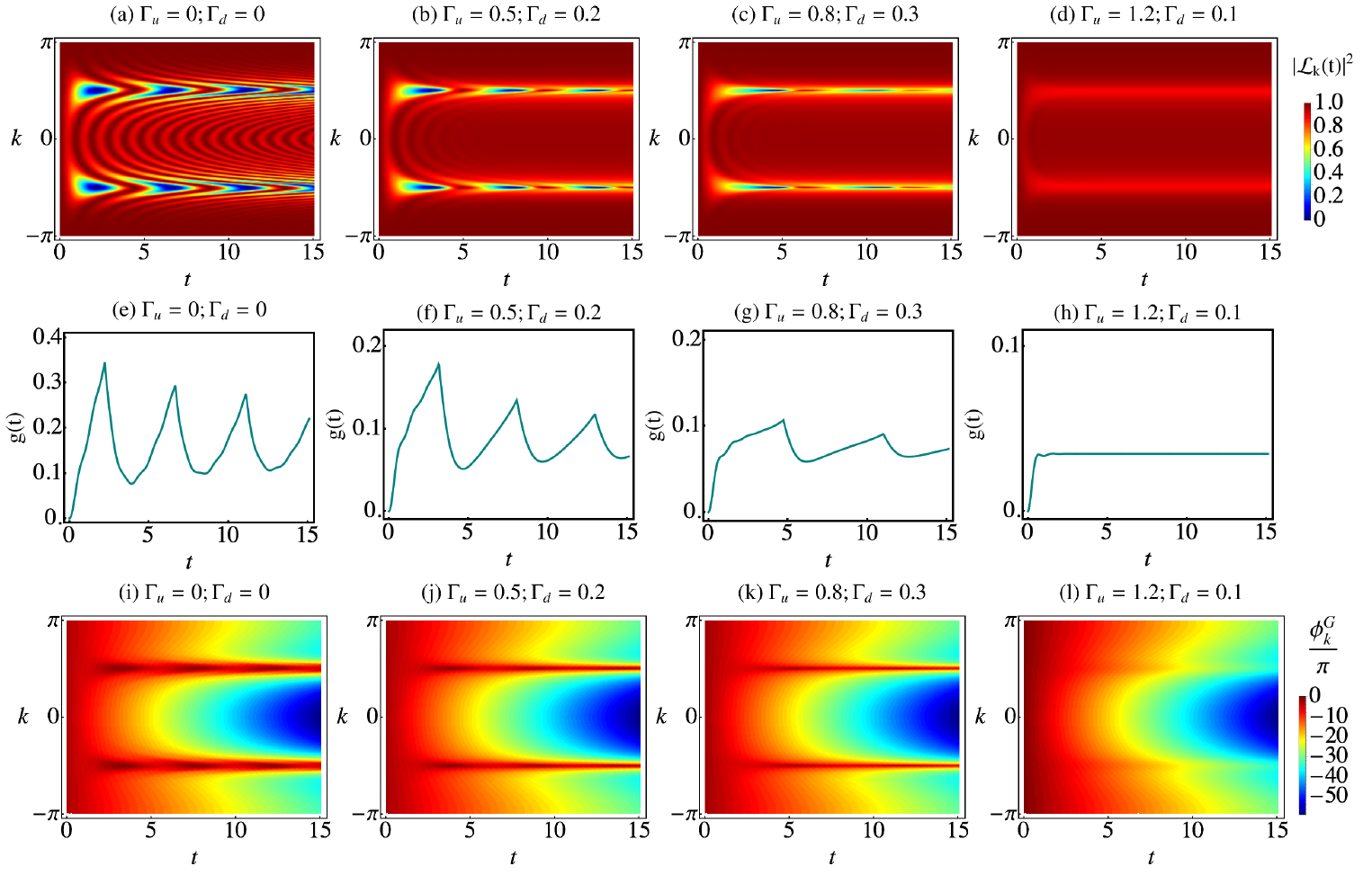}}
\centering
\caption{(Color online) The density plot of Loschmidt echo $|{\cal L}_{k}(t)|^{2}$
of periodically time driven extended XY model as a function of time $t$ and $k$,
for  (a) $\Gamma_{-}=0$, (b) $\Gamma_{-}=0.3$, (c) $\Gamma_{-}=0.5$, (d) $\Gamma_{-}=1.1$.
The dynamical free energy of the model versus time $t$ for (e) $\Gamma_{-}=0$,
(f)  $\Gamma_{-}=0.3$, (g) $\Gamma_{-}=0.5$, (h) $\Gamma_{-}=1.1$.
The density plot of geometric phase as a function of time and $k$
for  (i) $\Gamma_{-}=0$, (j) $\Gamma_{-}=0.3$, (k) $\Gamma_{-}=0.5$,
(l) $\Gamma_{-}=1.1$. In all plots we take $J_{1}=1, J_{2}=2\pi, h_{s}=3\pi,
\omega=6\pi$.}
\label{fig4}
\end{figure*}
%
and become gapless if
%
\bea
\bl
\label{eq27}
&
\Gamma_{-}
[
J_{2}\cos(k)+2h_{s}-\omega
]=0,\\
&
\frac{1}{4}
[
J_{2}\cos(k)+2h_{s}-\omega
]^{2}+
[J_{1}
\cos(\frac{k}{2})
]^{2}-{\Gamma_{-}}^{2}=0.
\quad\quad
\el
\eea
%

By solving these equations, we can get
%
\bea
\bl
\label{eq28}
&k^{\ast}=\arccos(\frac{\omega-2h_{s}}{J_{2}}),
\\
\label{eq29}
&\frac{2\Gamma_{-}^2}{J_{1}^2}-\frac{\omega-2h_{s}}{J_{2}}=1.
\el
\eea
%
The first term of  Eq.~(\ref{eq28}) implies a limitation $\omega-2h_{s}<\pm J_{2}$, and the second one 
defines an
exceptional points. Therefore, the system can be separated into three regions as shown in Fig.~\ref{fig3}.
In the region~(I), inside the exceptional closed curve, the energy gap $\Delta=|\epsilon_{K}^{+}-\epsilon_{k}^{-}|$
is purely real i.e., Im$[\Delta]=0$,
and Re$[\Delta]>0$. In this region $k^{\ast}=\arccos[(\omega-2h_{s})/J_{2}]$
and $\Delta=\sqrt{J_{1}^{2}(J_{2}+\omega-2h_{s})/(2J_{2})-\Gamma_{-}^{2}}$.
In the region~(II) the gap is pure imaginary, i.e., Im$[\Delta]\neq0$ and Re$[\Delta]=0$.
In this region we still have $k^{\ast}=\arccos[(\omega-2h_{s})/J_{2}]$ but
the non-Hermitian strength $\Gamma_{-}$ is large enough to be dominant, then
$\Delta={\it i}\sqrt{\Gamma_{-}^{2}-J_{1}^{2}(J_{2}+\omega-2h_{s})/(2J_{2})}$.
The region~(III) ($|\omega-2h_{s}|>J_{2}$) is characterised by the complex gap. In the other words, in the region~(III)
both real and imaginary parts of the gap is non-zero.

According to discussion in section \ref{GPXY}, the complex geometrical non-adiabatic phase for
the periodically time driven Floquet EXY model is also given by Eq.~(\ref{eq18}),
in which $h_{xy}(k)=J_{1}\cos(k/2)$ and $h_{z}(k)=J_{2}\cos(k)/2+h_{s}-{\it i}\Gamma_{-}$.
Then the real part of the complex geometrical non-adiabatic phase is given by
%
\bea
\no
{\rm Re}
[\beta]=
\left\{
\begin{array}{ll}
\pi, & \hbox{\small\text{Region(I)}} \\
\pi
[1+
\frac{
\Gamma_{-}}{\sqrt{\Gamma_{-}^{2}-J_{1}^{2}
(
J_{2}+\omega-2h_{s}
)
/(2J_{2})
}}],
 & \hbox{\small\text{Region(II)}}\\
\pi[1-f(k)], & \hbox{\small\text{Region(III)}}
\end{array}
\right.
\eea
%
where
$$f(k)=
\Big(
[2h_{s}+J_{2}\cos(k)-\omega]
/
(2{\rm Re}
[\Delta]
)
\Big)
-(\Gamma_{-}/
{\rm Im}
[\Delta]
)
.
$$
As seen the real part of the complex geometrical non-adiabatic phase shows singularity at
phase boundaries. It is necessary to mention that, all regions in Fig.~\ref{fig4}  are topologically  trivial and winding number
is zero.

\subsection{Pure state dynamical topological quantum phase transition\label{PDPTEXY}}

The Loschmidt amplitude for EXY model is calculated as
%
\begin{equation}
\bl
\label{eq30}
{\cal L}(k,t)=e^{-\Gamma_{+} t}e^{{\it i}\omega t}
\Big[
\frac{\cos(\Lambda t)+{\it i}
\frac{2h_{z}(k)-\omega}{2\Lambda}\sin(\Lambda t)}
{\sqrt{\langle\psi_{-}(k,t)|\psi_{-}(k,t)\rangle}}
\Big].
\el
\end{equation}
%
%
%
The DQPT occurs at the time instances at which at least one factor in LA becomes zero i.e., ${\cal L}_{k^{\ast}}(t^{\ast})=0$
which yields
%
\bea
\label{eq31}
t^{\ast}=\frac{-{\it i}}{2\Lambda}\ln
\Big[\frac{2h_{z}-\omega-2\Lambda}{2h_{z}-\omega+2\Lambda}\Big].
\eea
%
It straightforward to show that there are real solutions of $t^{\ast}$ only whenever
%
\bea
\label{eq32}
2h_{s}+J_{2}(\frac{2\Gamma_{-}^2}{J_{1}^2}-1)<\omega<2h_{s}+J_{2},
\eea
%
at quasi-momentum $k^{\ast}=\arccos[(\omega-2h_{s})/J_{2}]$ results
%
\bea
\label{eq33}
t^{\ast}=\frac{1}{2\Lambda}(2n+1)\pi+\frac{1}{\Lambda}\arctan(\frac{\Gamma_{-}}{\Lambda}).
\eea
%
According to Eq.~(\ref{eq32}) or Eq.~(\ref{eq29}),  DFDPTs exist in the range of driving frequency over which the eigenvalues of the time-independent effective Floquet non-Hermitian Hamiltonian are purely real but the system is not topological.
There is no critical momentum when $(\Gamma_{-}^2/J_{1}^2)-[(\omega-2h_{s})/J_{2}]>1$ and $t^{\ast}$ is always imaginary, resulting in no DFDPTs at any real time $t$. 
The lower bound of the driven frequency range across which DFDPTs occur is clearly reliant on dissipation, but the upper bound is independent of dissipation coupling.
Therefore, the range of driven frequency over which DFDPTs
occur shrinks to a single point $\omega=J_{2}+2h_{s}$ at $\Gamma_{-}=\pm J_{1}$.
It is worth noting that, in the absence of dissipation, FDPTs do not rely on the exchange coupling $J_1$,
 however, in the presence of dissipation, the DFDPT driven frequency range depends on $J_{1}$.

We present  the density plot of the Loschmidt echo $|{\cal L}(k,t)|^{2}$, the dynamical free energy $g(t)$,
and the density plot of geometric phase in Figs.~(\ref{fig4}) for different values of dissipation.
 Figs.~\ref{fig4}(a)-\ref{fig4}(c) show
 that when the eigenvalues of time-independent effective Floquet non-Hermitian EXY Hamiltonian $H_{F}$
are pure real, region~(I), there exist critical points $k^{\ast}$ and $t^{\ast}$, where ${\cal L}_{k^{\ast}}(t^{\ast})$ becomes zero.
 Contradiction, there is no such critical point out side of region~(I) [Figs.~\ref{fig4}(d)].
Moreover,  Figs.~\ref{fig4}(e)-\ref{fig4}(h) observe  DFDPTs  as  cusps in $g(t)$ for the driving frequency at which the system inters into
the region~(I), while $g(t)$ shows completely analytic, smooth behavior when the Hamiltonian parameters set out side
the region~(I).

The density plots of $\Phi^{G}_{k}$ are also plotted in Figs.~\ref{fig4}(i)-\ref{fig4}(l) for different values of Hamiltonian's parameters
inside and out side of the region~(I). As seen, the plots display singular changes at critical times $t^{\ast}$, and
at critical momentum $k^{\ast}$ when the system is in region~(I), while it shows smooth behavior for the case that the DFDPTs are absent.
This behaviour represents the topological aspects of DFDPTs, where the phase of the time-independent effective Floquet non-Hermitian
EXY Hamiltonian is not topological.

In is remarkable to mention that, in the absence of the dissipation, the lower bound of driven frequency range
in Eq.~(\ref{eq32}) and both lower and upper bounds of driven frequency range in Eq.~(\ref{eq22}) are
the critical points (gap closing) of the time-independent effective Floquet Hermitian Hamiltonians
in Eqs.~(\ref{eq25})~and~(\ref{eq7}). However, the upper bound of driven frequency range in Eq.~(\ref{eq22})
is not the critical point of the time-independent effective Floquet Hermitian Hamiltonian in Eqs.~(\ref{eq25}).
As a result, we may conclude that, in the absence of dissipation, only the gap closing (critical) points of the
time-independent effective Floquet Hermitian Hamiltonian are affected by dissipation.

\section{Conclusion}
We have investigated the dissipative Floquet dynamical phase transition in the periodically time
driven XY and extended XY models in the presence of the imaginary terms, which represent the physical gain and loss during
the interacting processes with the environment. We have shown that, the time-independent effective Floquet non-Hermitian Hamiltonians
reveal three regions with pure real eigenvalues (gap) where confined to exceptional points, pure imaginary gap and complex gap.
We have found that, the complex geometrical non-adiabatic phase can distinguish each regions of the system.
We have shown that the Floquet dynamical phase transitions still appearance in the presence
of the dissipation in the region where the time-independent effective Floquet non-Hermitian Hamiltonians exhibit real eigenvalues.
While the real gap region in the time-independent effective Floquet non-Hermitian XY Hamiltonian is topologically
nontrivial, its counterpart in the time-independent effective Floquet non-Hermitian extended XY Hamiltonian is topologically
trivial. In the other words, different from results obtained for the quenched case, existence of non-Hermitian topologically
nontrivial phase is not necessary condition for appearance of the dissipative Floquet dynamical phase transitions.
We have also shown that the range of driven frequency, over which the dissipative Floquet dynamical phase transitions occur,
narrows down by increasing the dissipation coupling and shrinks to a single point at the critical value of dissipation. Furthermore,
the topological characteristic aspect of the dissipative Floquet dynamical phase transitions in the real gap region is revealed by
quantization and jumps of the dynamical geometric phase.

\section*{Acknowledgments}
  A.~A. acknowledges the support
 of the Max Planck- POSTECH-Hsinchu Center for Complex Phase
 Materials, and financial support from the National Research
 Foundation (NRF) funded by the Ministry of Science of Korea (Grant
 No. 2016K1A4A01922028).


\appendix

\section{Spinless fermion transformation of the Floquet XY model\label{AA}}

The Hamiltonian, Eq.~(\ref{eq1}), can be diagonalized using the Jordan-Wigner transformation\cite{LIEB1961,Barouch1971,Jafari2011,Jafari2012}
%
\bea
\bl
&S^{+}_{n}= S^{x}_{n} + {\it i}S^{y}_{n}=\prod_{m=1}^{n-1}(1-2c_{m}^{\dagger}c_{m})c_{n}^{\dagger},\\
&S^{-}_{n}= S^{x}_{n} - {\it i}S^{y}_{n} = \prod_{m=1}^{n-1}c_{n}(1-2c_{m}^{\dagger}c_{m}),\\
&S^{z}_{n} = c_{n}^{\dagger}c_{n} -\frac{1}{2}.
\el
\eea
%
which transforms spins into fermion operators $c_{n}$, and $c^{\dagger}_{n}$.
Using the Fourier transform, the Hamiltonian of Eq.~(\ref{eq2}) can be written as the sum of $N/2$ non interacting terms
%
\bea
\label{eqA1}
{\cal H}(t) = \sum_{k>0} {\cal H}_{k}(t).
\eea
%
where this local Hamiltonian reads
%
\bea
\label{eqA2}
{\cal H}_{k}(t)&=&[J\cos(k)+h-{\it i}\Gamma_{-}] (c_{k}^{\dagger} c_{k}+c^{\dagger}_{-k} c_{-k})\\
\no
&-&{\it i}\gamma\sin(k)(e^{{\it i}\omega t}c_{k}c_{-k}-e^{-{\it i}\omega t}c_{k}^{\dagger}c_{-k}^{\dagger})-{\it i}\Gamma_{+},
\eea
%
where the wave number $k$ is equal to $k=(2p-1)\pi/N$ and $p$ runs from $1$ to $N/2$.
By defining the fermionic two-component spinor $C^{\dagger}=(c_{k}^{\dagger},~c_{-k})$ the Hamiltonian ${\cal H}(t)$
can be written as ${\cal H}_{k}(t)=C^{\dagger}\mathbb{H}_{k}(t)C-{\it i}\Gamma_{+}\mathbb{1}$, where
$\mathbb{H}_{k}(t)$ is given by Eq.~(\ref{eq6}).
We can get the eigenvalues and eigenvectors of Hamiltonian ${\cal H}_{k}(t)$ by solving the time-dependent Schr\"odinger equation:
%
\bea
\label{eqA5}
{\it i}\frac{d}{dt}|\psi(k,t)\rangle={\cal H}_{k}(t)|\psi(k,t)\rangle.
\eea
%
The exact solution to the Schr\"{o}dinger equation is found by going to the rotating frame given by the non-unitary transformation
$U(t)=U_{R}(t)U_{D}(t)$, with
%
\bea
\bl
\label{eqA6}
U_{R}(t)=\left(
           \begin{array}{cc}
             1 & 0 \\
             0 & e^{{\it i}\omega t} \\
           \end{array}
         \right),
         \quad
U_{D}(t)=\left(
           \begin{array}{cc}
             e^{-\Gamma_{+}t} & 0 \\
             0 & e^{-\Gamma_{+}t} \\
           \end{array}
         \right).
         \no
         \el
         \\
\eea
%
In the rotated frame the eigenstate is given by $|\varphi(k)\rangle=U^{-1}(t)|\psi(k,t)\rangle$.
Substituting the transformed eigenstate into Schr\"{o}dinger equation, we can obtain the time-independent
effective Flouquet non-Hermitian Hamiltonian:
%
\begin{equation}
\bl
\label{eqA7}
{\it i}\frac{d}{dt}|\varphi(k)\rangle=\Big[U^{-1}(t){\cal H}_{k}(t)U(t)-{\it i}U^{-1}(t)\frac{dU(t)}{dt}\Big]|\varphi(k)\rangle.
\el
\end{equation}
%
Under this unitary transformation the time-independent effective Flouquet non-Hermitian Hamiltonian $H_{F}$ is given by
Eq.~(\ref{eq7}).

\section{Complex geometrical non-adiabatic phase\label{AB}}

The instantaneous eigenstates of $I(t)$ and $I(t)^{\dag}$
are given as
%
\bea
\label{eqB4}
\bl
|\Psi_{+}\rangle
\no
=
\left(
\begin{array}{c}
\cos(
\frac{\alpha}{2}
) \\
\sin(
\frac{\alpha}{2}
)e^{{\it i}\omega t}
\\
\end{array}
\right),
\quad
%
|\Psi_{-}\rangle
=
\left(
\begin{array}{c}
-\sin(
\frac{\alpha}{2}
)e^{-{\it i}\omega t}
 \\
\cos(
\frac{\alpha}{2}
)\\
\end{array}
\right),
\el
\\
\eea
%
and
%
\bea
\bl
\label{eqB5}
\langle\Phi_{+}|
=&
\left(
\begin{array}{cc}
\cos(
\frac{\alpha}{2}
) & \quad \sin(
\frac{\alpha}{2}
)e^{-{\it i}\omega t} \\
\end{array}
\right),\\
\langle\Phi_{-}|
=&
\left(
\begin{array}{cc}
-\sin(
\frac{\alpha}{2})e^{{\it i}\omega t}
&
\quad
 \cos(\frac{\alpha}{2})\\
\end{array}
\right),
\el
\eea
%
respectively.

\bibliography{Refs}

\end{document}